\documentclass[preprint,aps,amsmath,nofootinbib,tightenlines,floatfix]{revtex4}
\usepackage{graphicx, epsfig}

\newcommand{\eeq}{\end{equation}}
\newcommand{\beq}{\begin{equation}}
\newcommand{\bea}{\begin{eqnarray}}
\newcommand{\eea}{\end{eqnarray}}

\begin{document}

\setlength{\unitlength}{1mm}

\title{The Inert Dark Matter}

\author{Ethan M. Dolle\footnote{edolle@physics.arizona.edu},
Shufang Su\footnote{shufang@physics.arizona.edu}}
\affiliation{Department of Physics, University of Arizona, Tucson, AZ 85721
}

\begin{abstract}
The lightest neutral scalar in the Inert Higgs Doublet Model is a natural candidate for WIMP dark matter.  In this paper, we analyzed the dark matter relic density in the Inert Higgs Doublet model.  Various theoretical and experimental constraints are taken into account.  We found that there are five distinctive regions that could provide the right amount of the relic density in the Universe.  Four out of those five regions have a light particle spectrum which could be studied at the Large Hadron Collider.

\end{abstract}

%\pacs{PACS Numbers: }

\maketitle

%\newpage

\section{Introduction}
\label{sec:intro}

About 20\% of the Universe is made of cold dark matter.  The origin of the dark matter,  however, still remains a mystery.  While the Standard Model (SM) of particle physics has been very successful in explaining the data from almost all of the particle physics experiments to date, none of the SM particles can be a good candidate for the dark matter.  Its existence provides unambiguous evidence for new physics beyond the Standard Model.

A Weakly Interacting Massive Particle (WIMP) is a promising candidate for the dark matter, given that the WIMP relic density is naturally around the observed value \cite{wmap}
\begin{equation}
\Omega h^2 = 0.112 \pm 0.009,
\end{equation}
for a WIMP mass around the electroweak scale $\sim 100$ GeV.   WIMPs also appear
naturally in many beyond the Standard Model scenarios.  One popular candidate is the lightest neutralino in supersymmetric models, which provides an example of spin-$1/2$ WIMP dark matter.  Spin-1 WIMP dark matter has been studied in the framework of extra dimension models, for example, the lightest Kaluza-Klein photon in Universal Extra Dimension models \cite{DMexD}.

Spin-0 dark matter has been studied \cite{SilveiraRK, McDonaldEX, BurgessYQ} in the framework of the SM plus an extra scalar.  The minimal model includes an extra real scalar gauge singlet $S$ that is charged under a $Z_2$ symmetry.  The only renormalizable couplings are the quartic self-coupling $\lambda_S S^4$ and the Higgs-$S$ coupling $\lambda S^2 H^\dagger H$.   It was shown that for a SM Higgs with mass in the range of 100 GeV $\leq m_h \leq 200$ GeV and dark matter with mass in the range of 10 GeV $\leq m_S \leq$ 100 GeV, viable regions of parameter space exists for the dark matter candidate $S$ to provide the right amount of   relic density in the Universe  \cite{BurgessYQ}.

Unlike the WIMP, the mass and the couplings of such a scalar dark matter particle often needs to be fine tuned to be consistent with the observed dark matter relic density.  WIMP-type scalar dark matter could appear in a simple extension of the SM with the addition of a Higgs ${\rm SU}(2)_L$ doublet.   Unlike the SM Higgs doublet, which couples to both fermion and gauge sectors, such an extra scalar doublet is {\it inert} in the sense that it couples to gauge bosons only.  Such inertness can be guaranteed by imposing a discrete $Z_2$ parity.  The lightest inert particle (LIP) is therefore stable.  If it is neutral, it can be a good WIMP dark matter candidate.

The Inert Higgs Doublet Model (IHDM)\footnote{Strictly speaking,  the extra ${\rm SU}(2)_L$ scalar  doublet is not a Higgs doublet since it does not obtain a vacuum expectation value.  Such model is also referred to as ``Inert Scalar Doublet Model"  in the literature. } was first proposed in the late 70's \cite{DeshpandeMa}.  It has received recent attention \cite{Barbieri} since it could be used to solve the naturalness problem in the Standard Model.   With a splitting in the masses of the neutral and charged components of the inert Higgs fields, the resulting positive contribution to the oblique $T$ parameter allows this model to  accommodate  a SM Higgs with a much heavier mass.  Such an inert Higgs sector  also appears in the recently proposed left-right Twin Higgs model \cite{twinhiggsleftright}.  In addition, an inert Higgs doublet (or doublets)  could be used to explain the small Majorana neutrino mass via the one-loop radiative seesaw mechanism \cite{Ma2006}, electroweak symmetry breaking \cite{HambyeVF}, grand unification \cite{LisantiEC} and leptogenesis \cite{MaFN}.

There have been some studies on the dark matter candidate in the IHDM.
Ref.~\cite{inerthiggsrelic} studied the dark matter relic density in the IHDM in certain regions of parameter space.   The right amount of relic density could be obtained for   dark matter with mass around
40 GeV $-$ 80 GeV or larger than 600 GeV.
Ref.~\cite{AgrawalXZ, AndreasHJ} studied the neutrino signatures from dark matter annihilation.  The continuous gamma ray spectrum from fragmentation and monochromatic gamma ray lines were studied in Ref.~\cite{inerthiggsrelic} and \cite{GustafssonPC} respectively.   Positron and antiproton signatures were studied in Ref.~\cite{NezriJD}.  Direct detection has been studied in \cite{Barbieri, inerthiggsrelic, MajumdarNT}.  There is also a collider analysis based on the LEP II limit  \cite{LundstromAI} as well as collider signatures of $SA$ associated production with $A\rightarrow S l^+l^-$  at the Large Hadron Collider (LHC) \cite{CaoRM}.

In this work, we performed a complete analysis of the dark matter relic density in the IHDM over the whole parameter space, taking into account various theoretical and experimental constraints.   The latest results of the collider constraints based on supersymmetric process $e^+e^-\rightarrow \chi_1^0\chi_2^0$ search at the LEP II are imposed \cite{LundstromAI}.
Unlike Ref.~\cite{inerthiggsrelic}, in which only a low SM Higgs mass $m_h=120$ GeV and 200 GeV are considered, our analysis also includes studies with a high SM Higgs mass $m_h=500$ GeV.  In Ref.~\cite{inerthiggsrelic}, the mass splittings
between $H^\pm$, $A$ and the dark matter candidate $S$, $(m_{H^\pm}-m_S, m_A-m_S)$,
 are fixed to be (50, 10) GeV and (10, 5) GeV  for the low and high dark matter mass regions, respectively.
We study  cases when the mass splittings between   $H^\pm$,  $A$  and   $S$ are small, in which   coannihilations  play an important role, as well as   cases when the mass splittings are large.
In regions that overlap with those analyzed in Ref.~\cite{inerthiggsrelic}, our results agree with the literature.
We identify additional regions of parameter space, in which the dark matter relic density is also consistent with the WMAP result but were overlooked previously.  We also present our results in the parameter spaces of physical Higgs masses and Higgs couplings,  which are convenient to use for   studies of collider phenomenology and dark matter detections.

The rest of the paper is organized as follows.  In Sec.~\ref{sec:model}, we briefly present the IHDM.  In Sec.~\ref{sec:constraints}, we discuss the theoretical and experimental constraints on the model parameter space.  In Sec.~\ref{sec:relic}, we present our results of the relic density analysis.   In Sec.~\ref{sec:conclusion}, we conclude.

\section{The Inert Higgs Doublet Model}
\label{sec:model}
The IHDM is an extension of the Higgs sector of the SM.  Besides  the usual Higgs doublet $H_1=H_{\rm SM}$, an additional Higgs doublet $H_2$ is introduced:
\begin{equation}
H_2=\left(
\begin{array}{c}
H^+\\(S+iA)/\sqrt{2}
\end{array}
\right),
\end{equation}
which is charged under ${\rm SU}(2)_L\times {\rm U}(1)_Y$ as $({\bf 2},1/2)$.   Unlike the SM Higgs boson, which couples to both SM gauge bosons and matter fermions, the extra Higgs doublet $H_2$ couples to the gauge sector only.  Such couplings can be guaranteed by imposing a $Z_2$ symmetry (sometimes also called matter parity) under which all particles except $H_2$ are even.  In particular, under the $Z_2$ symmetry:
\begin{equation}
H_1 \rightarrow H_1,\ \ \ H_2\rightarrow - H_2.
\end{equation}
While $H_1$ obtains a vacuum expectation value (VEV) $v/\sqrt{2}=174$ GeV  as in the SM, $H_2$ does not obtain a VEV: $\langle H_2 \rangle=0$.  The $Z_2$ symmetry is, therefore, not spontaneously broken.  The lightest particle in $H_2$ is stable and could be a good dark matter candidate.

The most general CP-conserving potential in the Higgs sector  that respects the $Z_2$ symmetry can be written as
\begin{equation}
V=\mu_1^2|H_1|^2+\mu_2^2|H_2|^2+
\lambda_1|H_1|^4+\lambda_2|H_2|^4+\lambda_3|H_1|^2|H_2|^2+
\lambda_4|H_1^\dagger H_2|^2+
\left[\frac{\lambda_5}{2}(H_1^\dagger H_2)^2+h.c.\right].
\label{eq:Higgspotential}
\end{equation}
Notice that the usual mixing term $\mu_{12}^2 H_1^\dagger H_2$ is forbidden by the $Z_2$ symmetry.
After electroweak symmetry breaking, three degrees of freedoms in $H_1$ are eaten by massive gauge bosons $W^\pm$ and $Z$.  We are left with one physical Higgs boson $h$, which resembles the SM Higgs boson, as well as four inert scalars: the CP even one $S$, the CP odd one $A$ and a pair of charged ones $H^\pm$.    The mass of $h$ is related to $\lambda_1$ via
\begin{equation}
m_h^2=-2\mu_1^2=2 \lambda_1 v^2.
\end{equation}
The masses of $S$, $A$ and $H^\pm$
are related to   parameters in the Higgs potential as
\begin{eqnarray}
m_{H^\pm}^2&=&\mu_2^2+\lambda_3 v^2/2,\\
m_{S}^2&=&\mu_2^2+(\lambda_3+\lambda_4+\lambda_5)v^2/2,\\
m_{A}^2&=&\mu_2^2+(\lambda_3+\lambda_4-\lambda_5)v^2/2.\\
\end{eqnarray}
We define the mass differences $\delta_1$ and $\delta_2$ as
\begin{equation}
\delta_1=m_{H^\pm}-m_S=-\frac{(\lambda_4+\lambda_5)v^2}{2(m_{H^\pm}+m_S)},\ \ \ \delta_2=m_{A}-m_S=-\frac{\lambda_5v^2}{(m_{A}+m_S)}.
\end{equation}
It is obvious that
$\lambda_4$ and $\lambda_5$ control the mass splitting between the charged and neutral CP even states, while $\lambda_5$ also controls the mass splitting between the CP odd and CP even states.
In our analysis below, we assume that the CP even scalar $S$ is the LIP, therefore the dark matter candidate, giving $\delta_{1,2}>0$.  The numerical results of the relic density analysis are similar if $A$ is the LIP dark matter.

The Higgs potential in Eq.~(\ref{eq:Higgspotential}) has seven parameters:
\begin{equation}
(\mu_1^2, \mu_2^2, \lambda_{1}, \lambda_2, \lambda_3, \lambda_4, \lambda_5).
\end{equation}
They could be replaced by the Higgs VEV $v$, physical Higgs masses, mass splittings,  and a sum of quartic couplings $\lambda_L=\lambda_3+\lambda_4+\lambda_5$ as
\begin{equation}
(v, m_h, m_S, \delta_1, \delta_2, \lambda_2, \lambda_L).
\end{equation}
In particular, $\lambda_L$ shows up in the couplings of $SSh$ and $SShh$, which is relevant for dark matter annihilation.  It is therefore convenient to pick $\lambda_L$ as a model parameter.  The quartic coupling $\lambda_2$ only shows up in self-couplings involving $S$, $A$ and $H^\pm$.  It does not play an important role for the dark matter analysis that we present below.

\section{Theoretical and experimental constraints}
\label{sec:constraints}

There are various experimental constraints on the IHDM from direct collider searches, indirect electroweak precision test and dark matter direct detections.

\begin{itemize}
\item{\bf $W$ and $Z$ decay widths}\\
Light $H^\pm$, $S$ and $A$ could lead to   deviations of $Z$ and $W$ decay widths from the SM value.  On the other hand, $\Gamma_{W,Z}$ have been measured precisely at the LEP and the Tevatron \cite{pdg}, which agree well with SM predictions.  Therefore regions in which the decay processes $W^\pm\rightarrow SH^{\pm}/AH^{\pm}$ and $Z\rightarrow H^+H^-/SA$ are kinematically allowed have already been excluded.  These translate into constraints for $m_S$ and $\delta_{1,2}$ as
\begin{eqnarray}
2m_S+\delta_1>m_W,&&2m_S+\delta_1+\delta_2>m_W,\nonumber \\
 2m_S+2\delta_1>m_Z, && 2m_S+\delta_2>m_Z.
\end{eqnarray}

\item{\bf Direct collider searches}\\
Light neutral and charged Higgses have been searched for at the LEP and the Tevatron.  Limits from conventional searches for $H^\pm$, $S$ and $A$, however, do not apply for  Higgses in the IHDM since those searches rely on decays of Higgses into fermion pairs and/or Higgs production via top quark decay.  Neutral and charged Higgses in the IHDM, on the other hand, do not couple to fermions.     In particular,
the charged Higgs has been searched for at the LEP and the
Tevatron \cite{chargedhiggslimitlep, chargedhiggslimitcdf}.
A lower mass bound of 74 $-$ 79 GeV at 95\% C.L. is obtained at the LEP \cite{chargedhiggslimitlep} considering $H^+\rightarrow c \bar{s}, \tau^+\nu$.
A more recent search at CDF \cite{chargedhiggslimitcdf} studied the charged Higgs produced in the top quark decay $t \rightarrow H^+ b$, with $H^+$ further decaying into a pair of quarks, leptons, or $W^+\phi$.   The bounds on the charged Higgs mass from those searches do not apply to Higgses in the IHDM  since Higgs-fermion-fermion couplings are absent.

Experimental signatures for $S$, $A$ and $H^\pm$, however, is similar to those of  neutralinos and charginos in the Minimal Supersymmetric Standard Model (MSSM), as $S$ appears as missing energy at colliders, similar to the lightest supersymmetric particle(LSP) in the MSSM.    In particular, searches of $e^+e^-\rightarrow \chi_1^0\chi_2^0$ at the LEP II can be interpreted as searches for  $e^+e^-\rightarrow SA$.
Therefore, the null result for   neutralino and chargino searches at the LEP II can be used to set limits on $m_{H^\pm}$, $m_A$ and $m_S$ in the IHDM.  A recent analysis \cite{LundstromAI} argued that a direct application of the upper limit on the LEP II $\chi_1^0\chi_2^0$ cross section to the IHDM is oversimplified due to the difference between the MSSM process $\chi_1^0\chi_2^0$ and IHDM process $SA$.   Based on DELPHI analyses of $e^+e^-\rightarrow \chi_1^0\chi_2^0$ with $\chi_2^0\rightarrow \chi_1^0 q\bar{q}, \mu^+\mu^-, e^+e^-$,  Ref.~\cite{LundstromAI} determined the efficiencies for the corresponding MSSM and IHDM processes after cuts using Monte-Carlo simulations.   The ratio of the efficiencies is then used to rescale the MSSM cross section upper limit and applied to the IHDM.   It is shown that regions satisfying
$m_S \lesssim 80$ GeV, $m_A\lesssim 100$ GeV and $\delta_2>8$ GeV are excluded by LEP II MSSM searches.  This limit is stronger than previous estimations by the direct application of the $\chi_1^0\chi_2^0$ cross section upper limit to the IHDM \cite{Barbieri, PierceUT}.
For $\delta_2<8$ GeV, however,  limits on scalar masses are much weaker,  due to the small mass splitting and the resulting soft jets and leptons in the final states.  For $\delta_2<8$ GeV, only the LEP I limit, $m_S+m_A>m_Z$, applies.

Similarly, searches of the supersymmetric channel $e^+e^-\rightarrow \chi_1^+\chi_1^-$ at the LEP II \cite{LEPchargino} can be used to set a bound on $m_{H^\pm}$.  Taking into account the cross section difference of scalars versus fermions, a limit of $m_{H^\pm} \gtrsim 70-90$ GeV can be derived from the LEP II chargino searches \cite{PierceUT}.

\item{\bf Electroweak precision test (EWPT)}\\
Electroweak precision measurements provide strong constraints for any new physics beyond the SM.   In the global electroweak precision fit to the SM, a light Higgs is preferred: $m_h=90^{+36}_{-27}$ GeV, with $m_h< 163 $ GeV at 95\% C.L. \cite{LEPEWWG}.  Oblique parameters \cite{PeskinSW} $S$, $T$ and $U$ are often used to parameterize radiative correction to gauge boson propagators.   A heavier SM Higgs contributes
positively to $ S$ while negatively to $  T$.  Extra Higgs bosons in the IHDM, namely $H^\pm$, $S$ and $A$,   also contribute to the $S$ and $T$ parameters,
which could possibly cancel the effect of a heavy SM Higgs boson \cite{Barbieri}.
Therefore, a heavy  SM Higgs can be accommodated in the IHDM.  In our analysis, we require the overall contribution  to the $S$ and $T$ parameters  from extra scalars in the IHDM to fall within the 68\% C.L. ellipse in $S-T$ plane.  Since  we considered possible large mass splittings $\delta_{1,2}$, where the approximate formulas for $\Delta S$ and $\Delta T$ in Ref.~\cite{Barbieri} might not be valid, we used the full expressions for the contributions to $\Delta S$ and $\Delta T$ for a Higgs doublet \cite{Barbieri} when imposing this constraint.

\begin{figure}[bht]
\begin{center}
\resizebox{3.in}{!}{\includegraphics*[83,230][510,562]{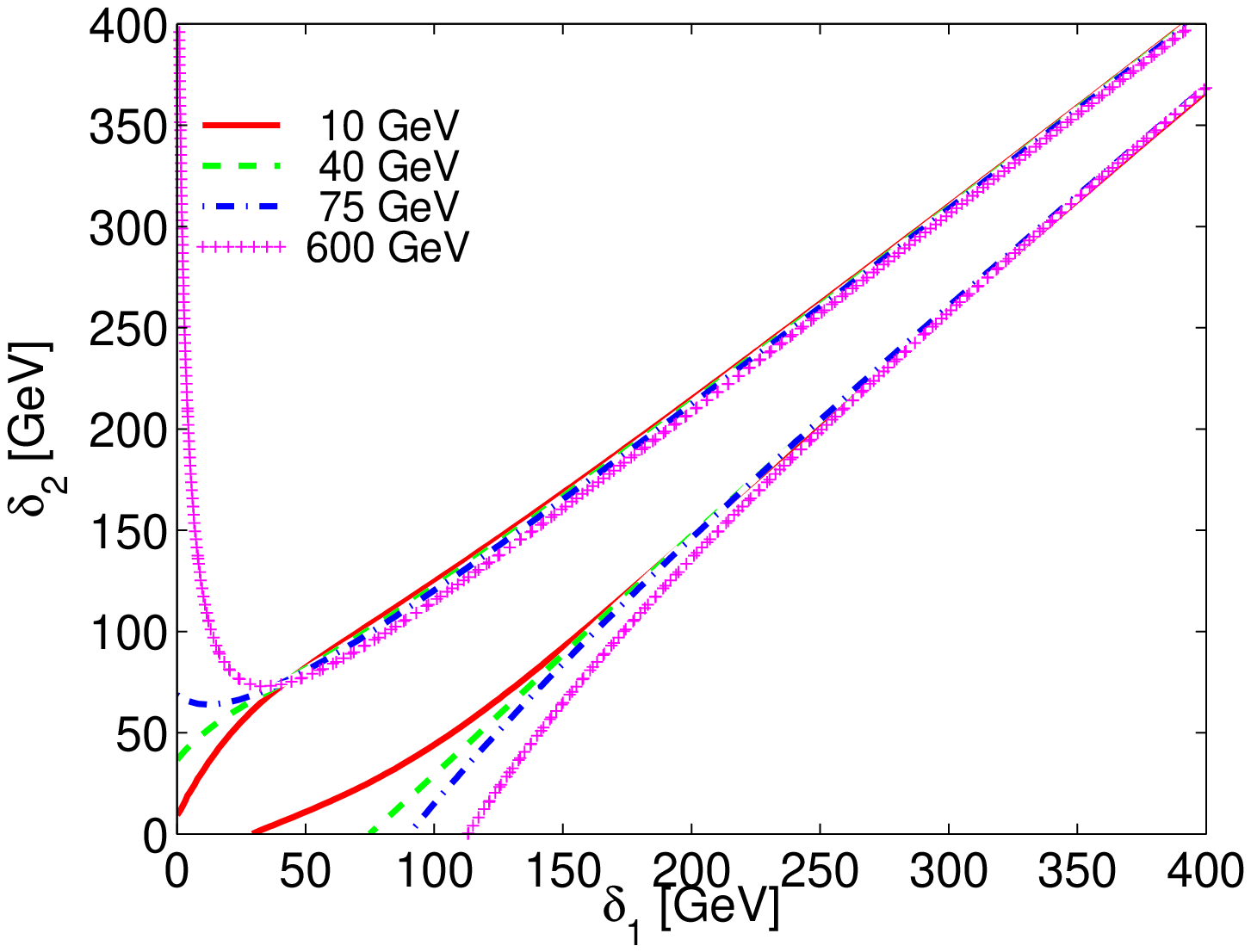}}
\resizebox{3.in}{!}{\includegraphics*[83,230][510,562]{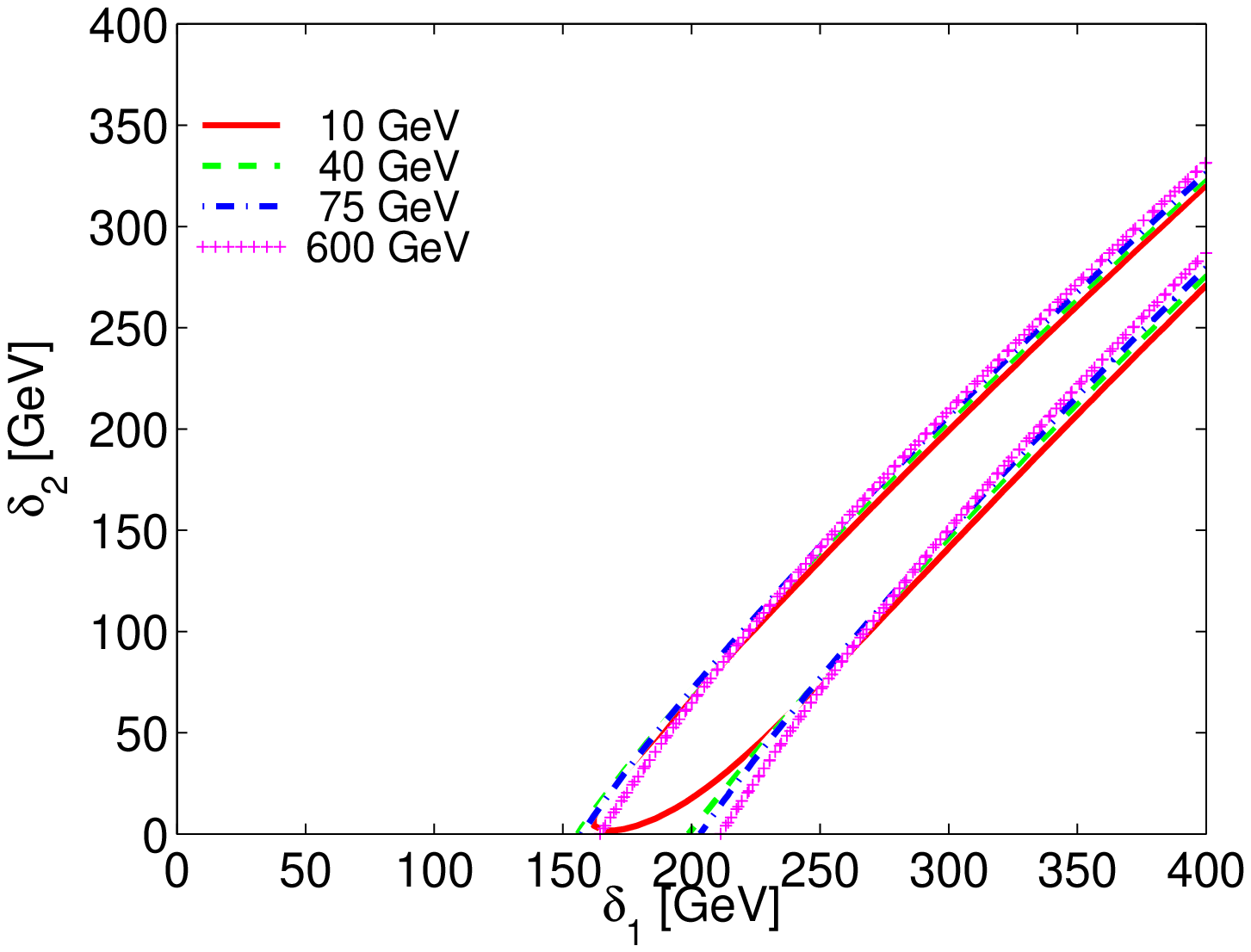}}
\caption{Plot of the 68\% C.L. allowed region in mass splittings $\delta_1$ and $\delta_2$ given the constraints on $S$ and $T$ from   precision electroweak measurements.  The value for the SM Higgs mass is set to be 120 GeV for the left plot and 500 GeV for the right plot.  $m_S$ is taken to be 10 GeV (solid curve), 40 GeV (dashed curve), 75 GeV (dash-dotted curve) and 600 GeV (dotted curve).  }
\label{fig:PEW}
\end{center}
\end{figure}

Fig.~\ref{fig:PEW} shows the allowed region for   mass splittings $\delta_1$ and $\delta_2$ given the constraints on the oblique parameters $S$ and $T$ from electroweak precision measurements, for several different choices of $m_S$.  The dependence on $m_S$ is small, especially for large $\delta_{1,2}$.  For $m_h=120$ GeV (left plot), the allowed region falls along the diagonal direction of $\delta_1 \sim \delta_2$.  The mass splittings $\delta_{1,2}$ could both be large  as long as these two mass splittings are close to each other.   For $m_h=500$ GeV (right plot), a large contribution from the inert Higgs doublet is needed to cancel the large  positive $S$ and  negative $T$ contribution from a heavy SM Higgs boson.  Therefore, the value for $\delta_1$ is constrained to be quite large $\delta_1 \gtrsim 150$ GeV, while the value for $\delta_2$ could still be as small as 0.

\item{\bf Dark matter detection}\\
Dark matter direct detection excluded spin-independent dark matter$-$nuclei scattering
cross section up to about $10^{-43}  {\rm cm}^{2} $ at 90\% C.L. \cite{XENON10, CDMS}, which is 7$-$8 orders of magnitude larger than dark matter $-$ nucleon scattering via $Z$ exchange.  This limit, however, can be easily avoided by introducing a small mass splitting between $S$ and $A$: $\delta_2 \gtrsim$ a few hundred KeV, given the typical kinetic energy
of the dark matter and the momentum transfer between the dark matter and the scattering nuclei in such scattering processes.

The spin-independent dark matter$-$nucleon scattering cross section via SM $h$-exchange is given by \cite{Barbieri}
\begin{equation}
\sigma=\frac{1}{4\pi}\left(\frac{m_{r}m_N}{m_S}\right)^2  \left(\frac{\lambda_Lf} {m_h^2}\right)^2
\end{equation}
where $m_N$ is the nucleon mass, $m_{r}$ is the reduced mass of the dark matter$-$nucleon pair, and $f$ is used to parameterize the nucleonic matrix element.
%, given by
%\begin{equation}
%\langle N|\sum m_q \bar{q}q|N\rangle =fm_N\langle N|N\rangle.
%\end{equation}
The typical range for $f$ is taken to be $0.14-0.66$ \cite{Tytgat}.   In our analysis, we take $f$ to be 0.14 to be on the safe side.
%We make no distinction between the coupling of $h$ with a proton or a neutron.
Latest results from XENON10, CDMS,  CRESST, CoGeNT and TEXONO are used in our analysis \cite{XENON10, CDMS, CDMSSUF, CRESST, CoGeNT, TEXONO}.  Portions of parameter space in small $m_S$, $m_h$ and large $|\lambda_L|$ region are excluded by the upper limits on the spin-independent dark matter$-$nucleon cross section.

The bounds from indirect dark matter detection (gamma rays, for example) is weak in the IHDM.  Moreover, those bounds are subject
to   large astrophysical uncertainties involved in those observations.
Therefore, we don't consider   constraints from indirect dark matter detections.

\end{itemize}

In addition, we impose the following theoretical constraints.
\begin{itemize}
\item{\bf Vacuum stability}\\
We require the vacuum stability of the Higgs potential at tree level, which leads to
\begin{eqnarray}
\lambda_{1,2}&>&0,\\
\lambda_3, \lambda_3+\lambda_4-|\lambda_5| &>& -2 \sqrt{\lambda_1 \lambda_2}.
\end{eqnarray}
\item{\bf Perturbativity}\\
%We require the quartic couplings $\lambda_i$ to be in the perturbativity region:
%$|\lambda_i|<4 \pi$.  Note that relatively large mass splitting $\delta_{1,2}$ would require $\lambda_4$ or $\lambda_5$ to be in the intermediate region $1<|\lambda_{4,5}|<4 \pi$.
We require that   corrections to the beta function of $\lambda_1$ from
non-SM quartic couplings $\lambda_{3,4,5}$ is less than the 50\% of the SM term $24 \lambda_1^2$ \cite{Barbieri}.  This amounts to the constraint:
\begin{equation}
\lambda_3^2+(\lambda_3+\lambda_4)^2+\lambda_5^2<12 \lambda_1^2.
\end{equation}
The evolution of the remaining quartic couplings does not lead to extra constraints.
In addition, we require the quartic coupling $\lambda_2$ to be in the perturbativity region:
\begin{equation}
\lambda_2<1.
\end{equation}

%\item{\bf Naturalness}\\
%We also impose naturalness constraints, which is to require  that   loop corrections  to the Higgs mass parameters not to exceed   tree level values for a cut off up to 1.5 TeV.  This leads to the constraints that  \cite{Barbieri}
%\begin{eqnarray}
%|2 \lambda_3+\lambda_4| &\lesssim& 9 \\
%\mu_2 &\gtrsim& (1, 2.5 \sqrt{\lambda_2}, \sqrt{|2 \lambda_3 + \lambda_4|})
%{\rm 120 \  GeV}
%\end{eqnarray}
%{\bf did you take into account these two constraints?}
%{\bf check Naturalness constraints.}

\end{itemize}

\section{Relic Density Analysis}
\label{sec:relic}
We analyzed the relic density in the IHDM using the program MicrOMEGAs \cite{micromegas}.  This program solves the Boltzmann equation numerically, using  the program CalcHEP \cite{calchep} to calculate all of the relevant cross sections.    When the mass splittings between the dark matter candidate and other particles are small, coannihilation effects are also included.

In our analysis, we fixed the SM Higgs mass to be 120 GeV (preferred by the EWPT in the SM) or 500 GeV (preferred by naturalness, and could be consistent with the EWPT with large splittings between $H^\pm$, $S$ and $A$.).   We fixed $\lambda_2=0.1$  since it does not enter  into the dark matter relic density analysis.
We studied both the cases of small mass splittings and large mass splittings.  When $\delta_1$($\delta_2$) is small, coannihilation between $S$ and $H^\pm$($A$) is important.   Part of our study overlaps with the analysis in Ref.~\cite{inerthiggsrelic}, which, only considered small Higgs masses $m_h=120$ GeV and 200 GeV with fixed mass splittings $(\delta_1, \delta_2)=(50,10)$ GeV for the low mass region and   $(\delta_1, \delta_2)=(10,5)$ GeV for the high mass region.
Results in Ref.~\cite{inerthiggsrelic} were presented in $\mu_2-m_S$ plane.
We present our results in $\lambda_L-m_S$ plane instead, which is more transparent since $m_S$ and $\lambda_L$ are the two relevant parameters for the dark matter analysis,  as well as studies of dark matter direct and indirect detections.   In particular, the value of $\lambda_L$ is closely related to the neutrino and gamma ray flux from dark matter annihilation, and the cross section for dark matter$-$nucleon scattering.
Larger $\lambda_L$ typically leads to enhanced flux and increased dark matter$-$nucleon scattering cross sections, and the inert Higgs dark matter has a better chance to be detected at future dark matter detection experiments.

We discuss below in detail two mass regions of $m_{{S}}$ that could provide the amount of dark matter relic density consistent with the WMAP result at   the 3$\sigma$ level:  (A) low mass region where $m_{{S}}<$ 100 GeV, and  (B) high mass region where 400 GeV $ < m_{{S}} <$ around a TeV.

\subsection{Low mass region}
\label{sec:lowmass}

\begin{figure}[bht]
\begin{center}
\resizebox{3.1in}{!}{\includegraphics*[83,230][510,562]{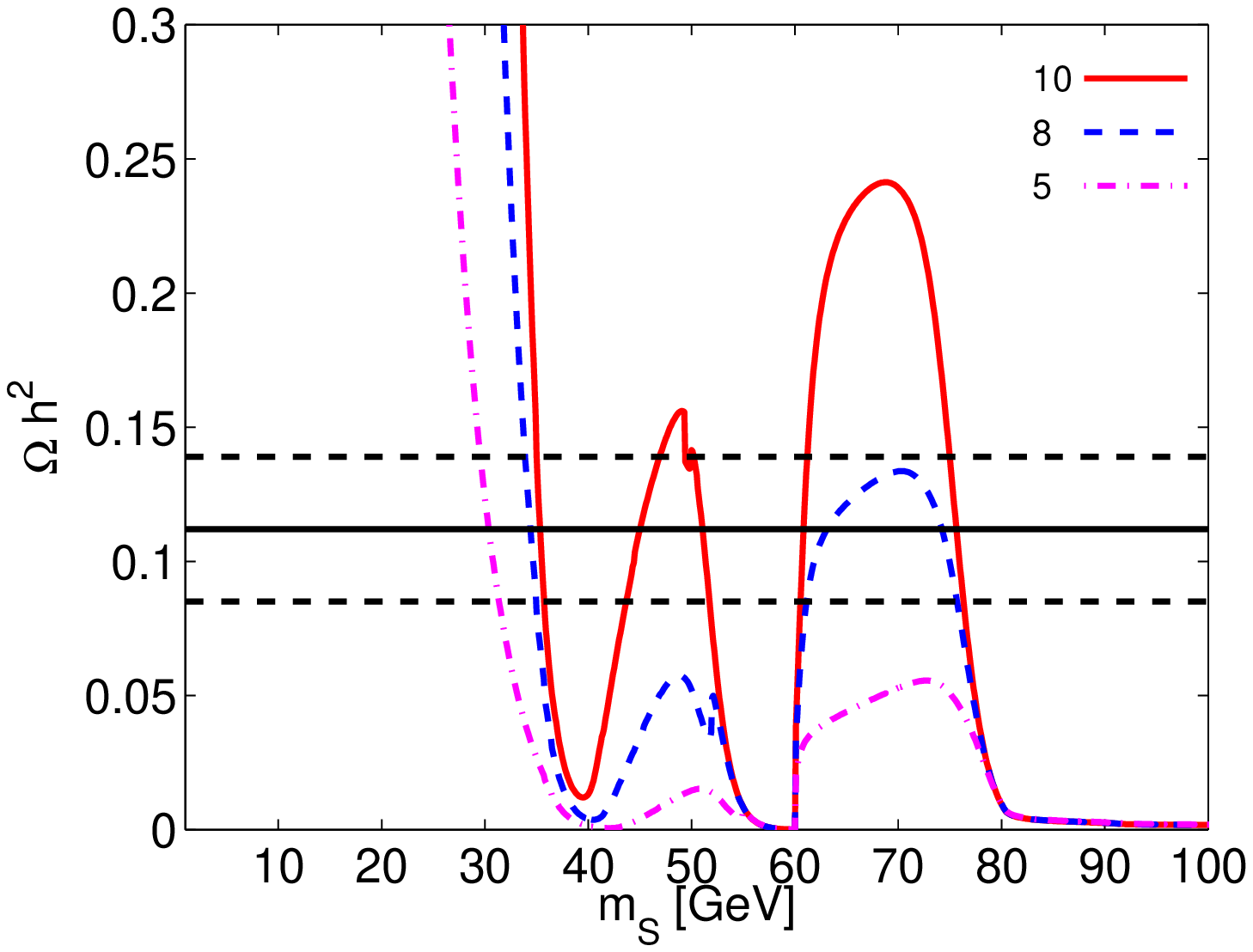}}
\resizebox{3.1in}{!}{\includegraphics*[83,230][510,562]{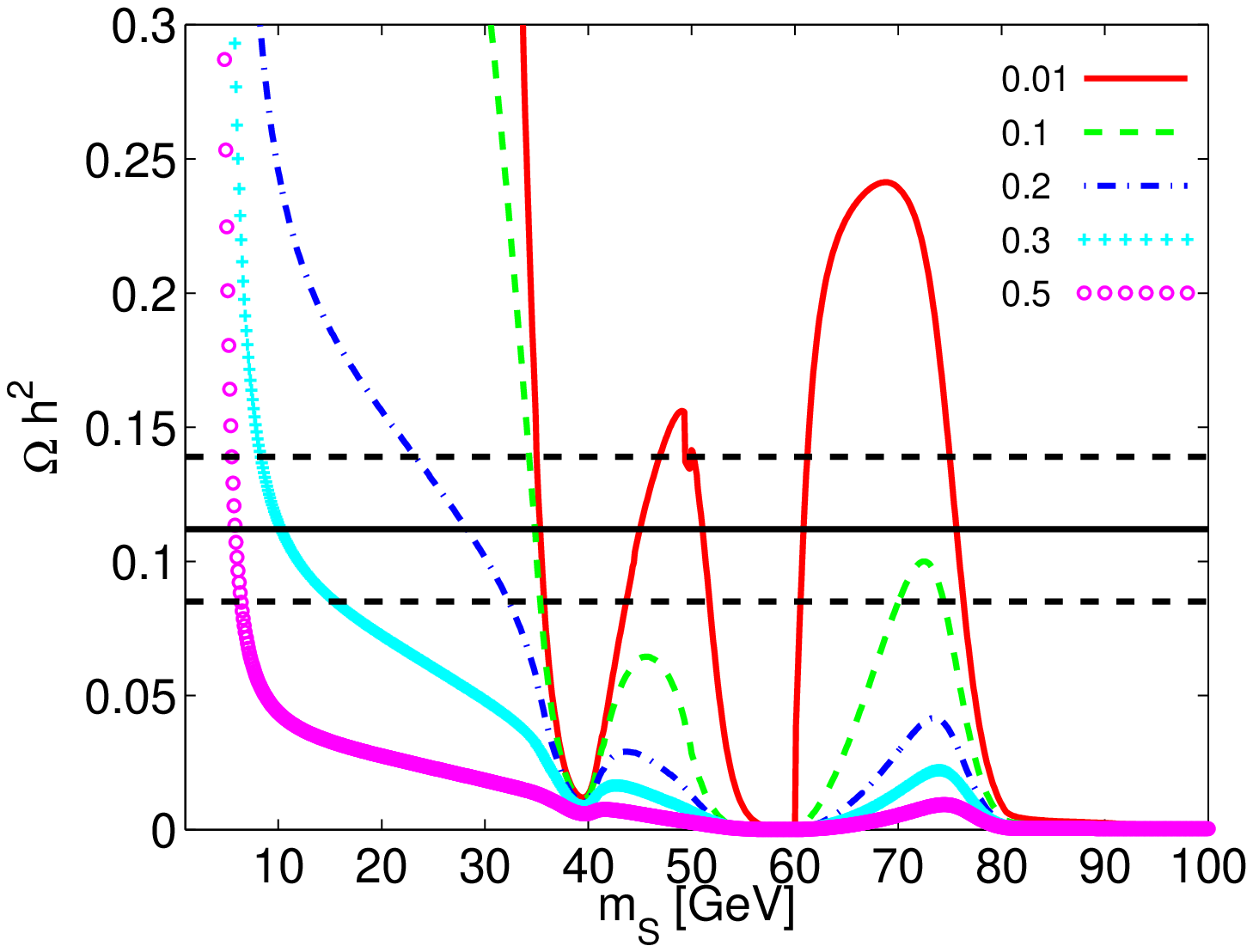}}
 \caption{Dependence of the relic density on $m_S$ for $m_h$=120 GeV and $\delta_1=50$ GeV.
In the left plot, $\delta_2$ is chosen to be 10 GeV (solid line), 8 GeV (dashed line) and 5 GeV (dash-dotted line) while $\lambda_L$ is fixed to be  0.01.
In the right plot, $\lambda_L$ is chosen to be 0.01 (solid line), 0.1(dashed line), 0.2 (dash-dotted line), 0.3 ``+" line), 0.5 (``o" line)  while $\delta_2$ is fixed to be  10 GeV.
The horizontal band indicates the 3$\sigma$ region that is consistent with relic density measurement from WMAP: $0.085<\Omega h^2<0.139$.
}
\label{fig:low_120_relic}
\end{center}
\end{figure}

To illustrate the dependence of the relic density $\Omega h^2$ on $m_S$, $\lambda_L$ and mass splittings $\delta_{1,2}$, we present in Fig.~\ref{fig:low_120_relic}   curves of $\Omega h^2$ vs. $m_S$ for various choices of $\delta_2$ (left plot) and $\lambda_L$ (right plot). The horizontal band indicates the 3$\sigma$ region that is consistent with relic density measurement from WMAP: $0.085<\Omega h^2<0.139$.
A light SM Higgs mass $m_h=120$ GeV is used in both plots.

The red curve in the left plot of Fig.~\ref{fig:low_120_relic} corresponds to $(\delta_1, \delta_2)=(50,10)$ GeV. Therefore, coannihilation between $S$ and $A$ is important while coannihilation between $S$ and $H^\pm$  is not.
For small $m_S$, $SS\rightarrow b\bar{b}$ via SM $h$ exchange dominates, with cross section proportional to $\lambda_L^2$.    The cross section is typically small due to the small bottom Yukawa coupling, which leads to relic density too big that overcloses the Universe.
When $m_S$ gets larger, $SA\rightarrow q \bar{q}$ via $Z$ exchange becomes more and more important.
The  relic density enters  the WMAP $3\sigma$ region for $m_S$ around 35 GeV.  The coannihilation cross section maximizes at the $Z$-pole: $m_S+m_A\sim m_Z$, corresponding to the dip around $m_S\sim 40$ GeV.  The coannihilation cross section decreases when $m_S$ increases   away from the $Z$-pole region, which makes the relic density falls back to the allowed region.  As $m_S$ gets larger, $SS\rightarrow b\bar{b}$ annihilation via $h$ starts to dominate and the relic density enters the $3\sigma$ region again.  For $m_S\sim m_h/2$, $SS$ annihilation hits the $h$-pole,  corresponding to the second dip around $m_S \sim 60$ GeV.  The annihilation cross section gets smaller once $m_S$ leaves the $h$-pole region.  When  $m_S \gtrsim 70$ GeV, $SS\rightarrow WW$ dominates.  The annihilation cross section quickly increases and the relic density  drops below the WMAP observed value.

The dashed and dot-dashed curves in the left plot of Fig.~\ref{fig:low_120_relic} shows the relic density dependence for $\delta_2$=8 GeV and 5 GeV respectively.   Coannihilation effects get stronger for smaller mass splittings.  Therefore, for most of the $m_S$ region  between 40 $-$ 60 GeV, the coannihilation cross section is too large and the relic density is too small.

Curves in the right plot of Fig.~\ref{fig:low_120_relic} correspond to $\lambda_L$=0.01, 0.1, 0.2, 0.3 and 0.5, respectively, while $(\delta_1, \delta_2)$ is fixed to be $(50,10)$ GeV.  Similar $Z$-pole and $h$-pole features appear.  The relic density is smaller for larger $\lambda_L$,
since $SS$ annihilation via $h$-exchange is increased due to the increased $SSh$ coupling.

\begin{figure}[bht]
\begin{center}
\resizebox{3.in}{!}{\includegraphics*[83,230][510,562]{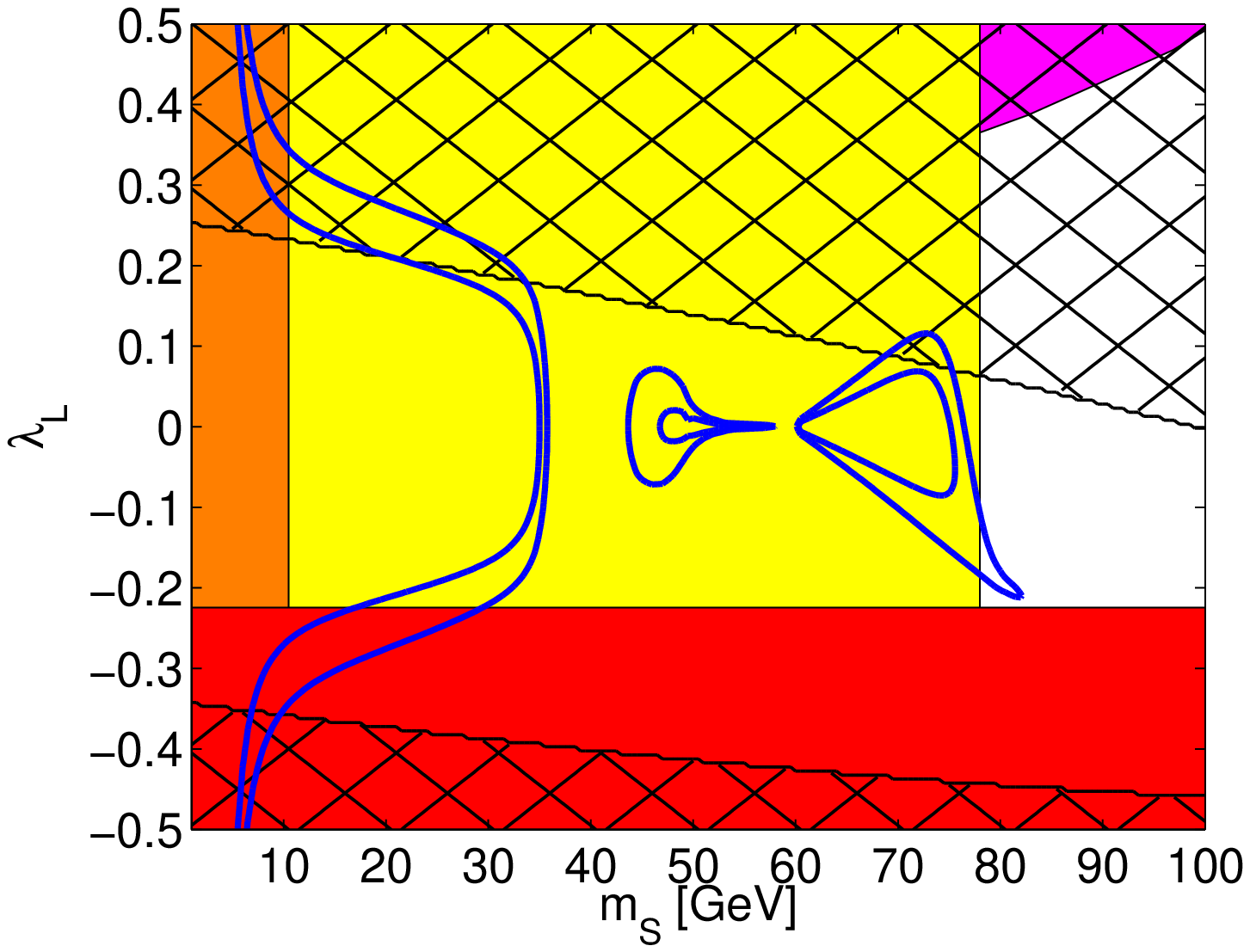}}
\resizebox{3.in}{!}{\includegraphics*[83,230][510,562]{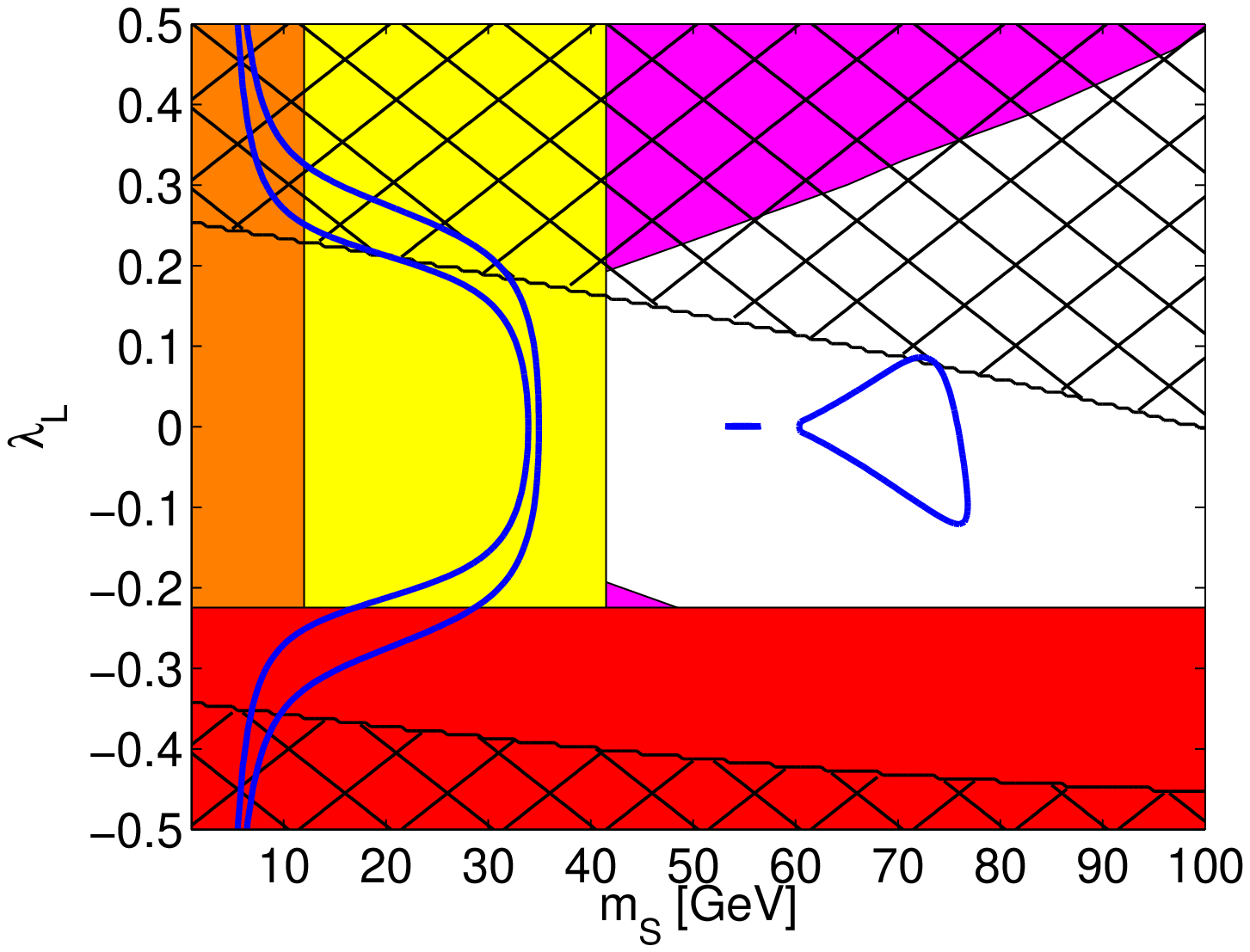}}
\caption{WMAP 3$\sigma$ allowed region (enclosed by blue curves) in  $m_S-\lambda_L$ plane for $m_h$=120 GeV.  The mass splittings are chosen to be $(\delta_1, \delta_2)=(50,10)$ GeV (left plot) and (50,8) GeV (right plot). Shaded regions are excluded either by LEP I+II searches (yellow, light shade), electroweak precision constraints (orange, medium shade), dark matter direct detection (purple, medium-dark shade), vacuum stability (red, dark shade along bottom), and  perturbativity (hatched region). }
\label{fig:low_120_1}
\end{center}
\end{figure}

Fig.~\ref{fig:low_120_1} shows the WMAP 3$\sigma$ relic density allowed region (enclosed by two blue curves) in the $m_S-\lambda_L$ plane for $m_S<100$ GeV with $m_h=120$ GeV for $(\delta_1, \delta_2)=$(50, 10) GeV (left plot) and (50, 8) GeV (right plot).   Shaded regions are excluded  by various theoretical and experimental constraints, as described in Sec.~\ref{sec:constraints}.

For  $(\delta_1, \delta_2)=$(50, 10) GeV (left plot),
the gap around $m_S\sim 40$ GeV corresponds to the $Z$-pole.
The gap around $m_S\sim 60$ GeV corresponds to the $h$-pole.
The LEP constraint (yellow, light shade  region)  is very strong due to the strong constraints on $m_A$ and $m_S$ when $\delta_2>8$ GeV.  The precision electroweak constraints (orange, medium shade region) is weak since $\delta_1>\delta_2$ is slightly preferred by the fit to the $S-T$ contour. Given all the constraints, only a small region around $m_S\sim 80$ GeV survives.  The value for $\lambda_L$ for the allowed region, however, could be as large as $-0.2$.    Such a large value of $\lambda_L$ would be important for generating a large signal in the indirect detection of dark matter.

The LEP constraints on $m_S$ and $m_A$, however, are weakened for small mass splitting $\delta_2 \lesssim 8$ GeV.   For such a small mass splitting, $m_S$ as low as around 40 GeV is still allowed.   In the right plot of Fig.~\ref{fig:low_120_1}, the allowed parameter space is given for   $(\delta_1, \delta_2)=$(50, 8) GeV.  In most of the $m_S$ region between 40 GeV and 60 GeV, the relic density for the dark matter is too small due to the large coannihilation $SA$ cross section.  However, there is a viable region for 60 GeV $<m_S<80$ GeV   with $\lambda_L$ in the region of $\pm0.1$.

\begin{figure}[bht]
\begin{center}
\resizebox{3.1in}{!}{\includegraphics*[83,230][510,562]{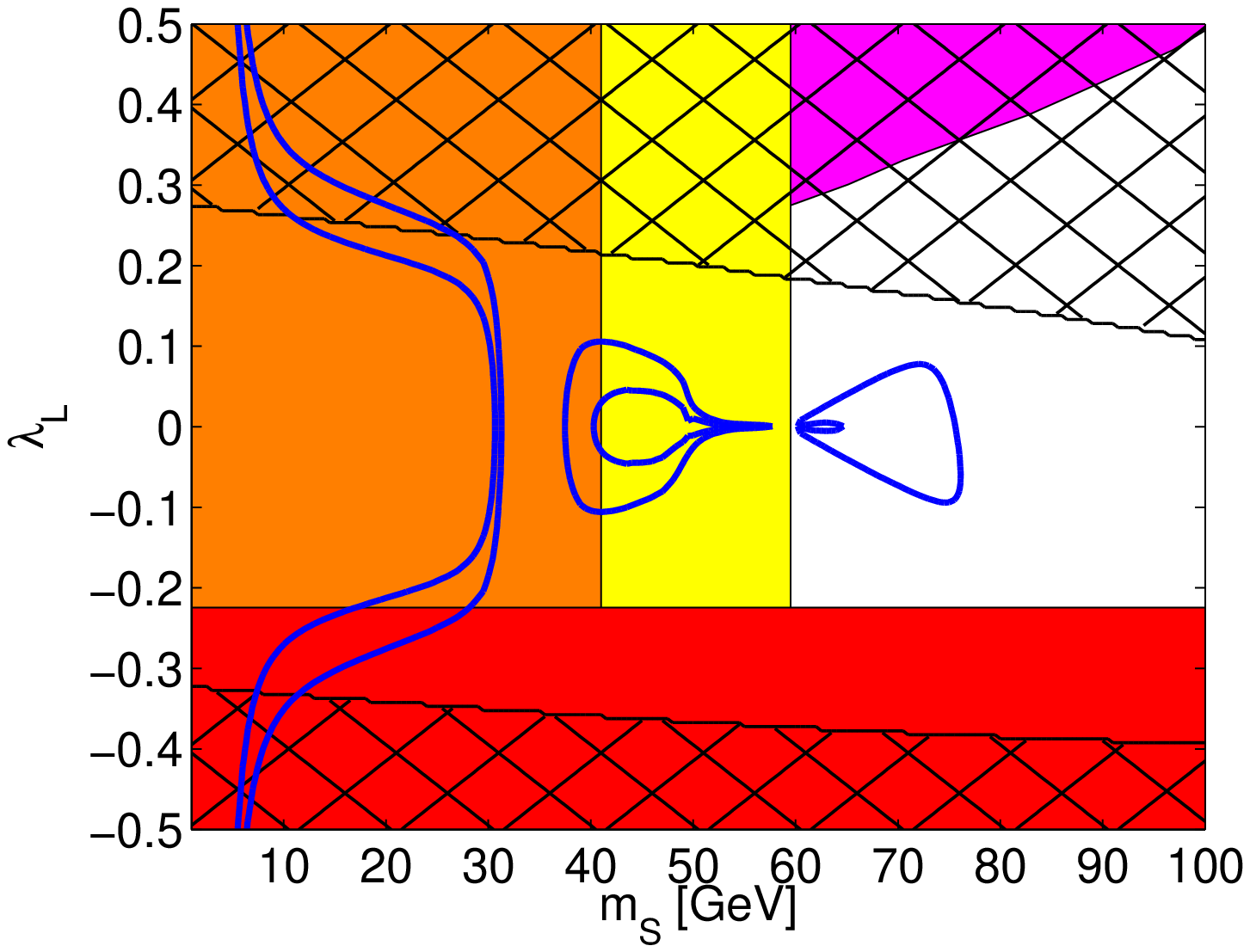}}
\resizebox{3.1in}{!}{\includegraphics*[83,230][510,562]{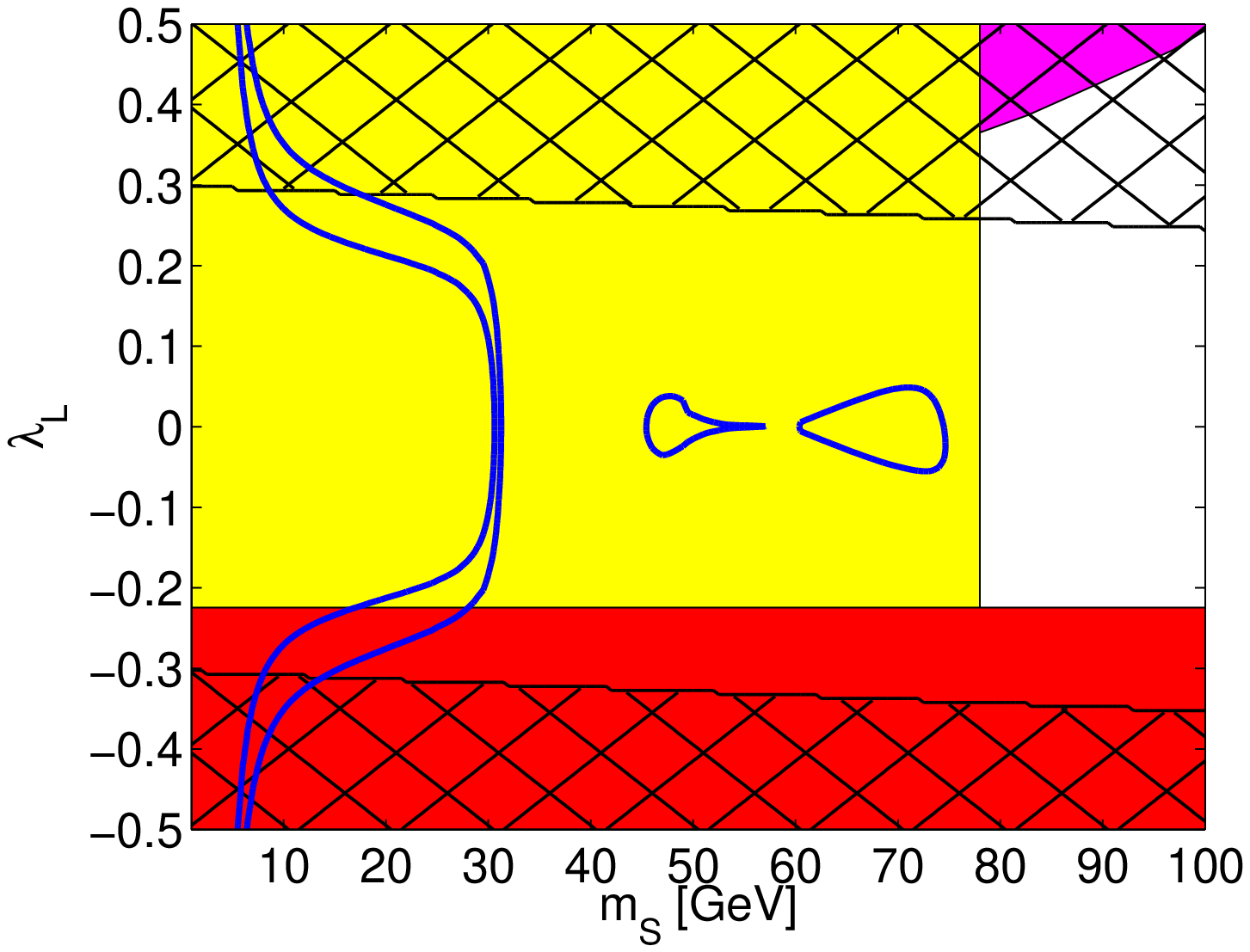}}
\caption{WMAP 3$\sigma$ allowed region (enclosed by blue curves) in  $m_S-\lambda_L$ plane for $m_h$=120 GeV.  The mass splittings are chosen to be $(\delta_1, \delta_2)=(10,50)$ GeV (left plot), (10,10) GeV (right plot).  Shaded regions are excluded by various theoretical and experimental constraints;  see caption of Fig.~\ref{fig:low_120_1} for explanation.    }
\label{fig:low_120_2}
\end{center}
\end{figure}

The left plot of Fig.~\ref{fig:low_120_2} shows the allowed relic density region for $(\delta_1, \delta_2)=$(10, 50) GeV and $m_h=120$ GeV.  The dark matter annihilation behaves similarly to the (50,10) case, only that the previous $SA$  coannihilation via $Z$  is replaced by  $SH^\pm$ coannihilation via $W^\pm$.
The LEP constraints (yellow, light shade region) becomes weaker comparing to $(\delta_1,\delta_2)=(50, 10)$ GeV case, due to the weaker constraints on $m_{H^\pm}$.  The precision electroweak constraints (orange, medium shade region), however,  becomes stronger.
Given all the theoretical and experimental constraints, a region with 60 GeV $<m_S<$ 80 GeV and $-0.1<\lambda_L<0.1$ survives.

When $\delta_1$ and $\delta_2$ are both small, the coannihilations between $S$, $A$ and $H^\pm$  are relevant and the allowed 3$\sigma$ regions shrink due to the enhanced coannihilation cross sections.  For ($\delta_1, \delta_2)$=(10, 10) GeV, there is no allowed region that survives given all the theoretical and experimental constraints, which is shown in the right plot of Fig.~\ref{fig:low_120_2}.  For $\delta_2\lesssim 8$ GeV, although the LEP II constraints is weaker, the WMAP 3$\sigma$ region  for $m_S>40$ GeV completely disappears due to   strong coannihilation effects.

\begin{figure}[tbh]
\begin{center}
\resizebox{3.in}{!}{\includegraphics*[83,230][510,562]{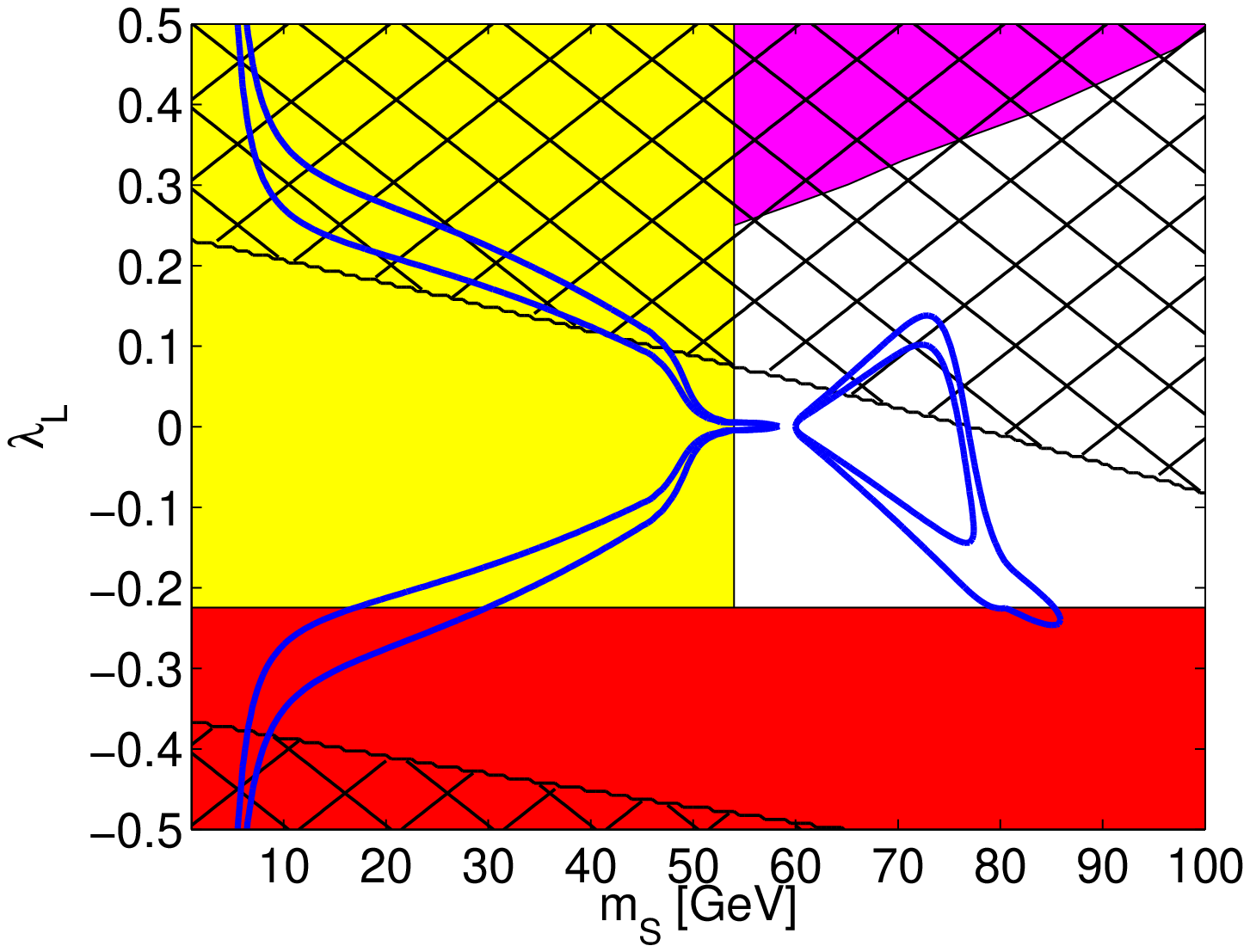}}
\resizebox{3.in}{!}{\includegraphics*[83,230][510,562]{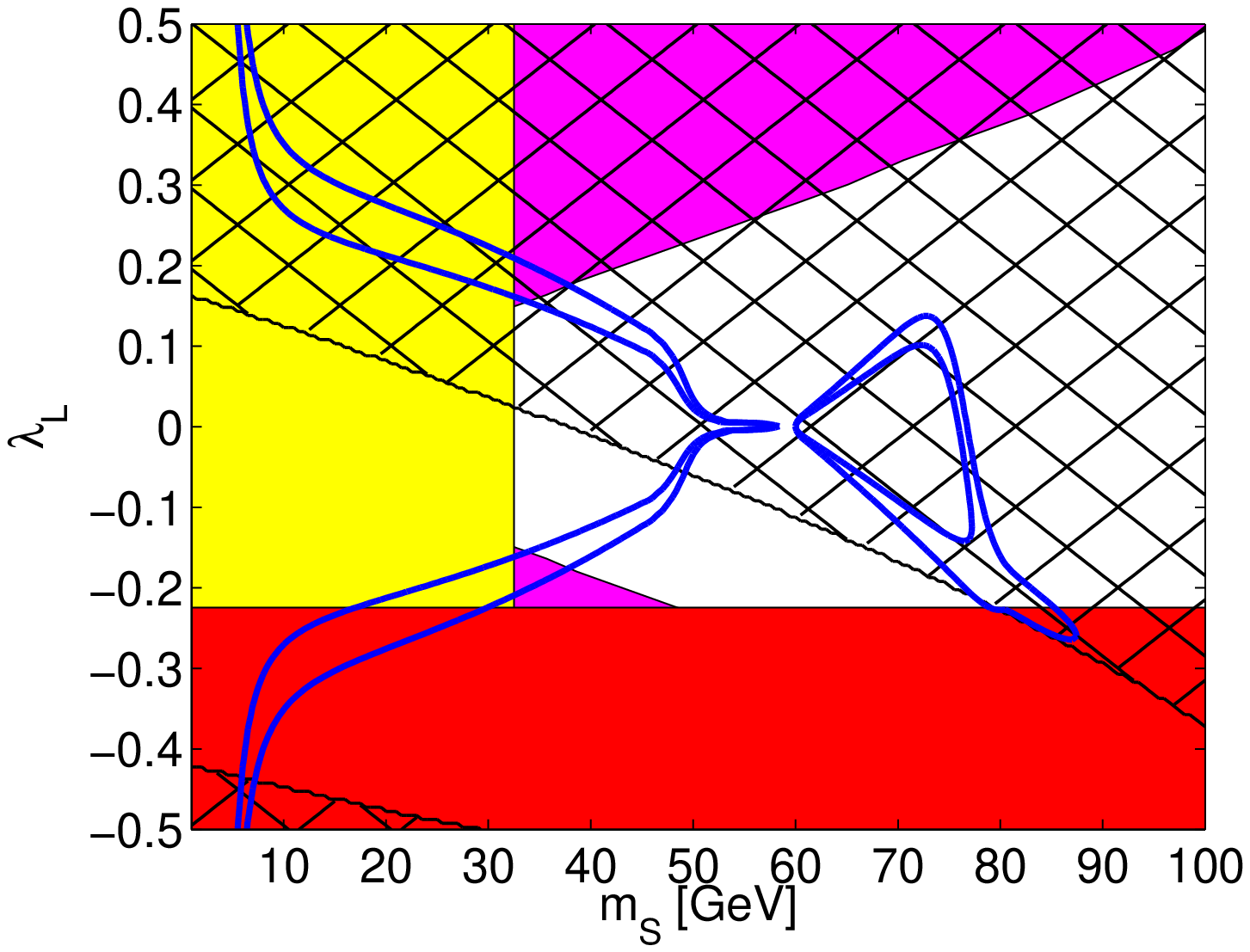}}
\caption{WMAP 3$\sigma$ allowed region (enclosed by blue curves) in  $m_S-\lambda_L$ plane for $m_h$=120 GeV.  The mass splittings are chosen to be $(\delta_1, \delta_2)=(50,50)$ GeV (left plot), (70,70) GeV (right plot).  Shaded regions are excluded by various theoretical and experimental constraints;  see caption of Fig.~\ref{fig:low_120_1} for explanation.  }
\label{fig:low_120_3}
\end{center}
\end{figure}

Fig.~\ref{fig:low_120_3} shows the allowed relic density region for $m_h=120$ GeV when both $\delta_{1,2}$ are large: $(\delta_1, \delta_2)=(50,50)$ GeV (left plot) and $(\delta_1, \delta_2)=(70,70)$ GeV (right plot).
The gaps in the low mass region corresponding to the $Z$-pole or $W$-pole disappear, since the mass splitting is too large for the coannihilation process to be important.
The $h$-pole, however, still survives due to the low value of $m_h=120$ GeV that we pick.  The LEP I+II constraints  (yellow, light shade region) are weaker because of  the large value of $\delta_2$.  The perturbativity constraints (hatched region), however, are stronger, since larger mass splittings $\delta_{1,2}$ correspond to larger values for $\lambda_{4,5}$.

For splitting $(\delta_1, \delta_2)=(50,50)$ GeV, a region with 55 GeV $<m_S<$ 90 GeV and $-0.2 < \lambda_L < 0$ is consistent with the WMAP 3$\sigma$ region.
When the mass splittings get larger, the perturbativity constraint gets stronger (the hatched region shifts to the left) while the LEP II constraint gets weaker (the yellow, light shade region also shifts to the left).  For $(\delta_1, \delta_2)=(70,70)$ GeV, a region with
30 GeV $<m_S<$ 50 GeV and $-0.15 < \lambda_L < -0.05$ opens up.  For $(\delta_1, \delta_2)=(90,90)$ GeV
the perturbativity constraint is so strong that no mass window survives all of the constraints.

\begin{figure}[bht]
\begin{center}
\resizebox{3.in}{!}{\includegraphics*[83,230][510,562]{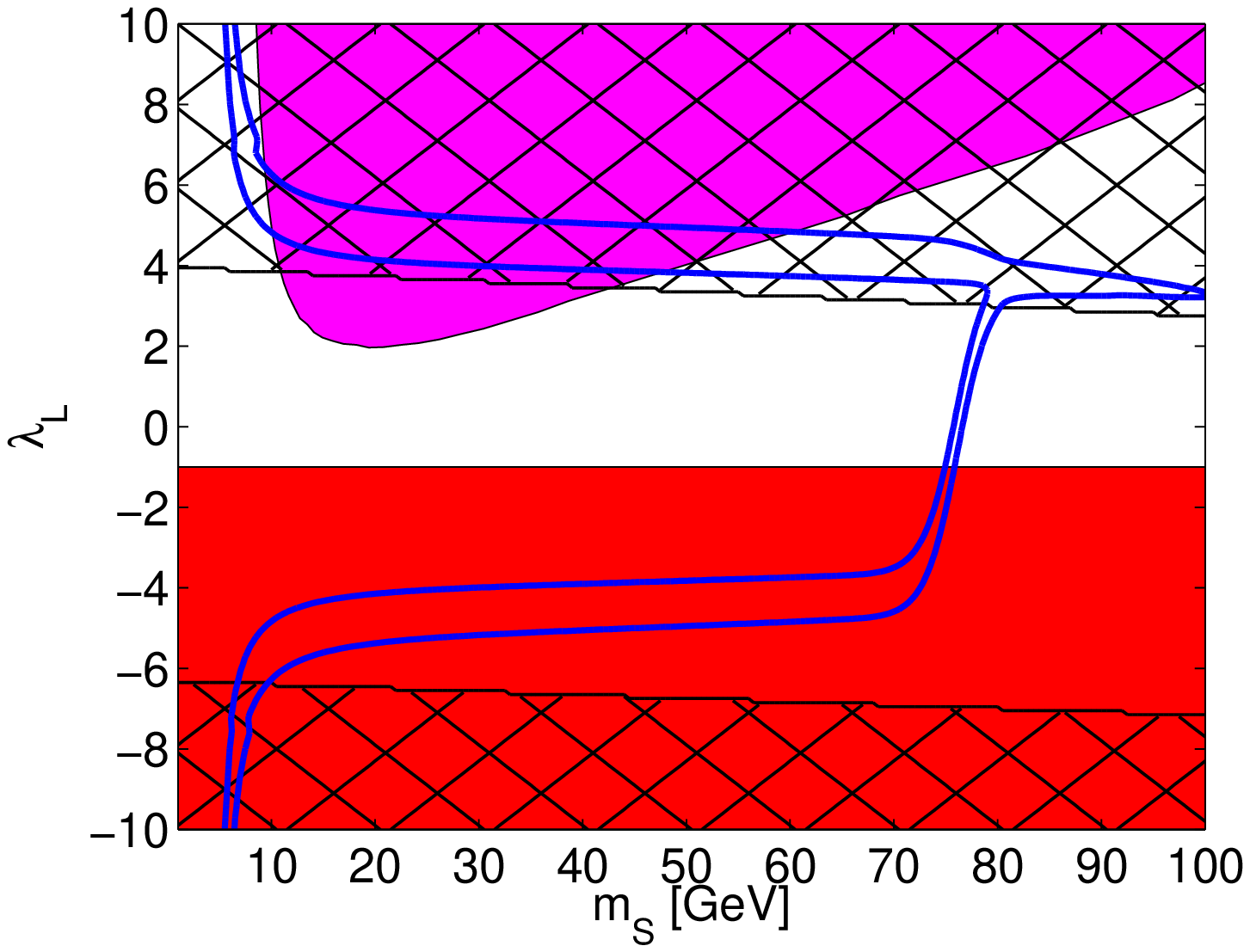}}
\resizebox{3.in}{!}{\includegraphics*[83,230][510,562]{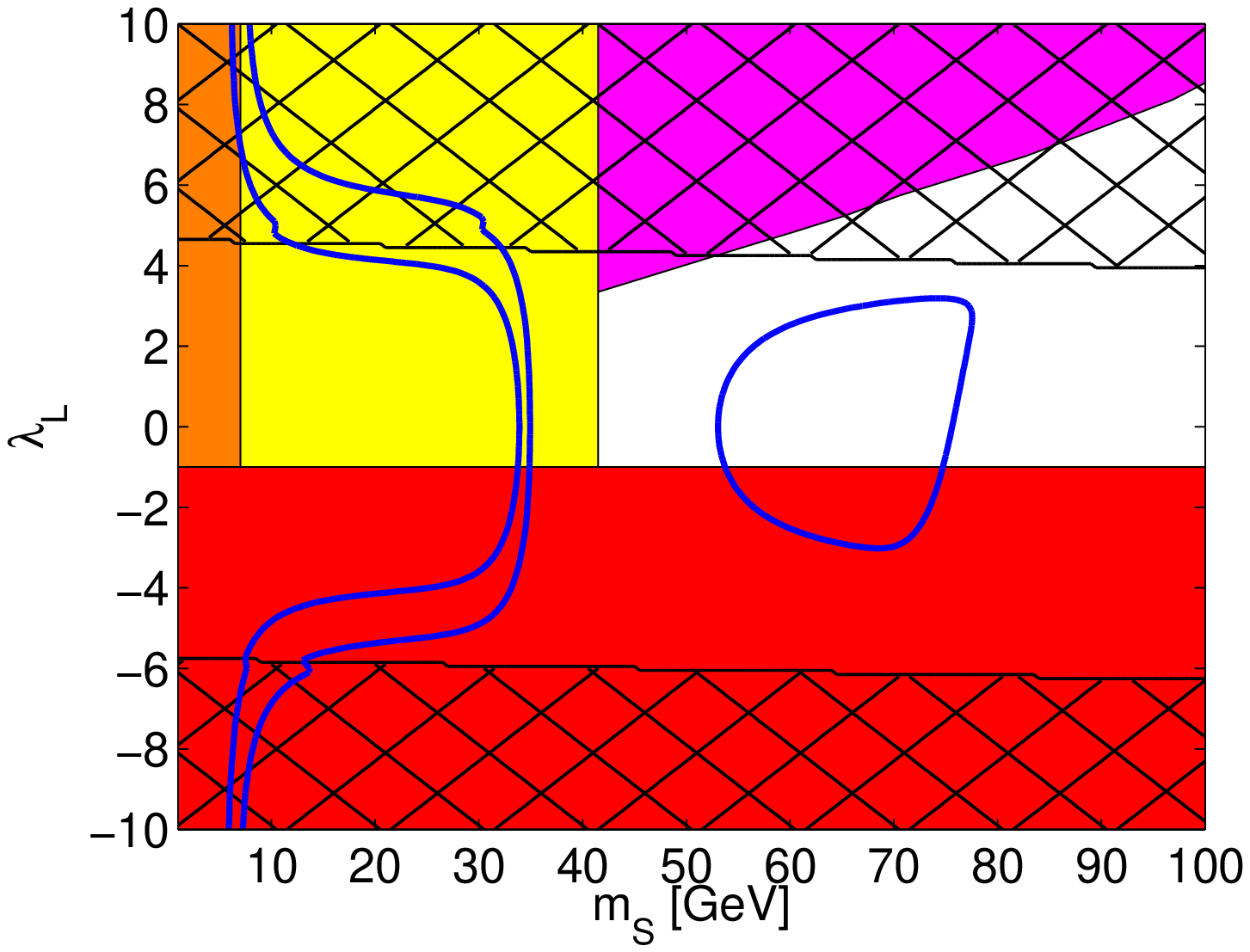}}
\caption{WMAP 3$\sigma$ allowed region (enclosed by blue curves) in  $m_S-\lambda_L$ plane for $m_h$=500 GeV.  The mass splittings are chosen to be $(\delta_1, \delta_2)=(250,110)$ GeV (left plot), (180,8) GeV (right plot).  Shaded regions are excluded by various theoretical and experimental constraints;  see caption of Fig.~\ref{fig:low_120_1} for explanation.  }
\label{fig:low_500}
\end{center}
\end{figure}

%All of the 120 GeV Higgs plots
%satisfy naturalness (except for the small region of the 100,100 splitting
%plot, which contains benchmark point 1).

Fig.~\ref{fig:low_500} shows the allowed relic density region for a high value of the SM Higgs mass $m_h=500$ GeV.
Electroweak precision constraints require a large value for $\delta_1$ (as shown in Fig.~\ref{fig:PEW}) in order to compensate for the $\Delta S>0$ and $\Delta T <0$ contributions from a heavy SM Higgs.
The left plot of Fig.~\ref{fig:low_500} shows the allowed relic density region for
$(\delta_1, \delta_2)=(250,110)$ GeV.  No pole regions appear due to the large mass splittings and large $m_h$.
For $m_S\lesssim 75$ GeV, $SS\rightarrow b\bar{b}$ dominates, which corresponds to the nearly horizontal band of the 3$\sigma$ WMAP region.  Once $m_S\gtrsim$ 75 GeV, $SS\rightarrow WW$ opens up, which leads to the nearly vertical band of the 3$\sigma$ region.
Dark matter with mass around 75 GeV is allowed given all of the constraints.
Note that the perturbativity bounds are much weaker due to the large SM Higgs mass (therefore, large $\lambda_1$).  On the other hand, to obtain the right relic density, large $\lambda_L$ is needed to compensate   the suppression of the annihilation cross section by the large Higgs mass $m_h$.
Although $\lambda_L$ could be as large as 3,  the indirect detection in this region is not promising due to large suppression of the annihilation cross section by $m_h^2$.

The numerical results do not change much for smaller values of $\delta_2$.
A $Z$-pole coannihilation region appears  for $\delta_2$ around 10 GeV.
The right plot of  Fig.~\ref{fig:low_500}  shows the allowed relic density region for $(\delta_1, \delta_2)=(180,8)$ GeV.
The gap around $m_S\sim$ 40 GeV is due to the $Z$-pole.
The LEP II constraints on $m_S$ and $m_A$ do not apply, due to the small mass splitting.  This allows  a relatively large $m_S$ region around $50-80$ GeV to remain open, with   $-1<\lambda_L<  3$.
This region, however, shrinks for smaller $\delta_2$ due to the stronger coannihilation effects.
Therefore, no WMAP 3$\sigma$ region survives for $\delta_2$ less than about 6 GeV.

\subsection{High mass region}
\label{sec:highmass}

\begin{figure}[bht]
\begin{center}
\resizebox{3.in}{!}{\includegraphics*[83,230][510,562]{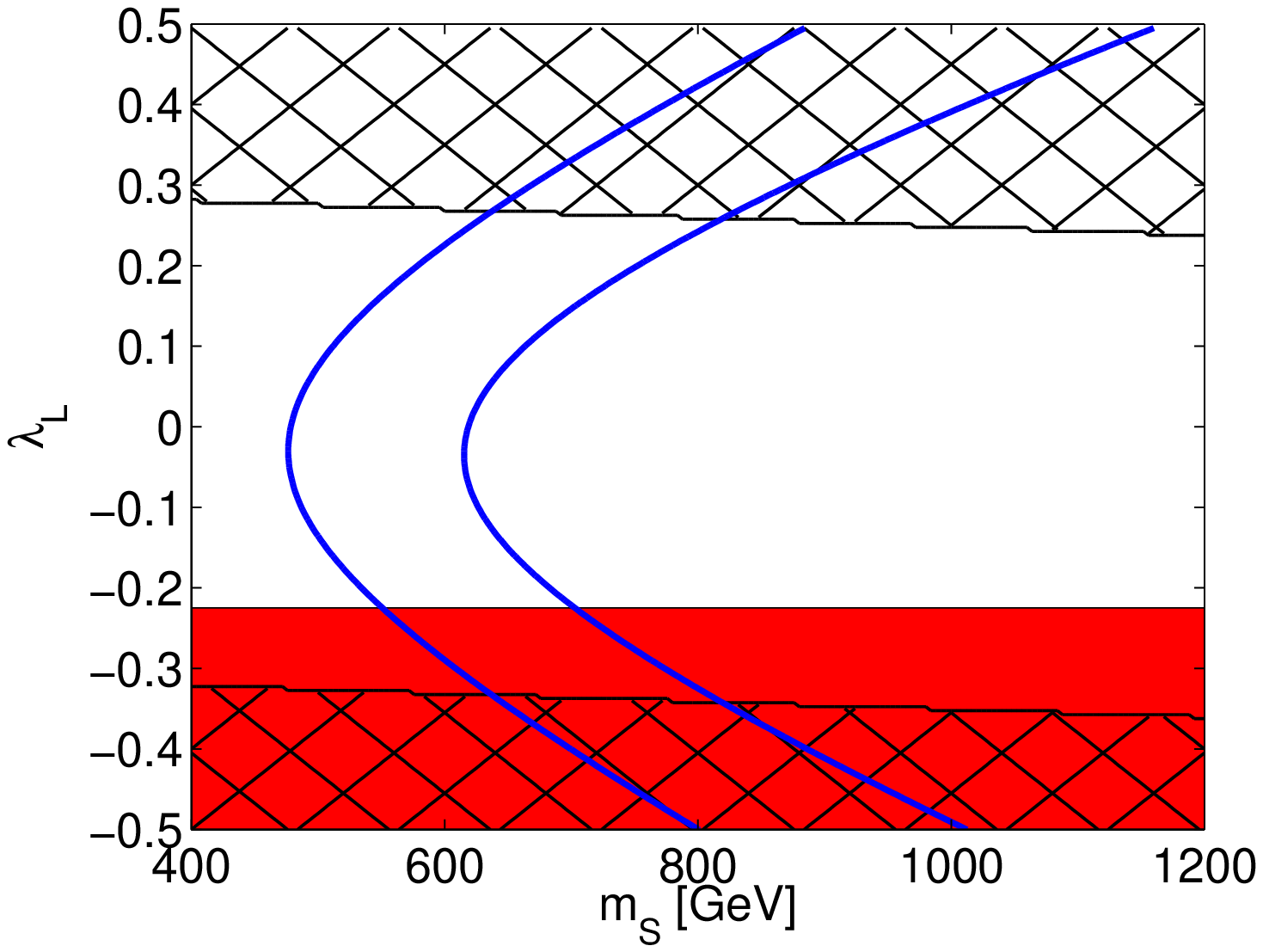}}
\resizebox{3.in}{!}{\includegraphics*[83,230][510,562]{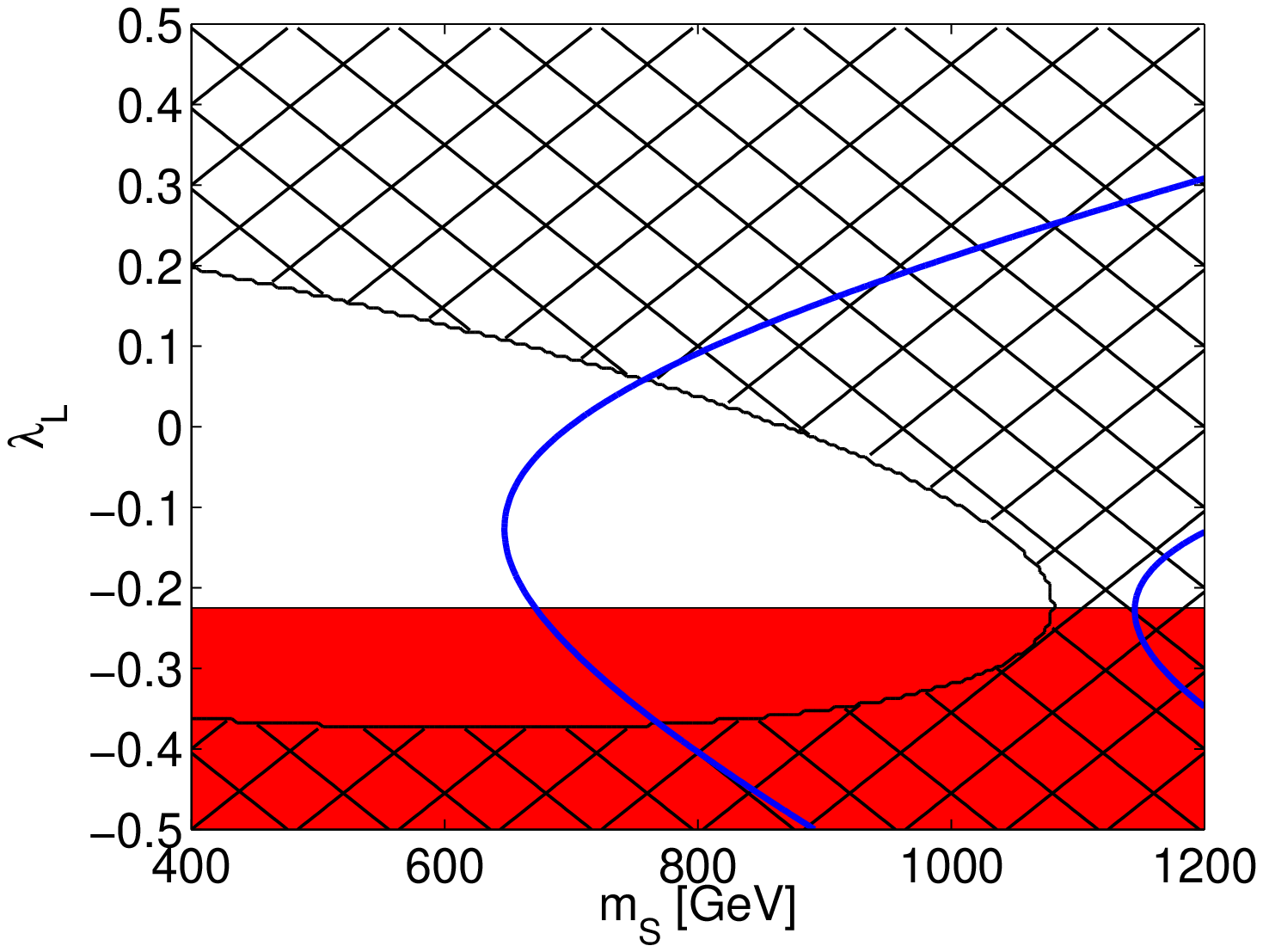}}
\caption{WMAP 3$\sigma$ allowed region (enclosed by blue curves) in  $m_S-\lambda_L$ plane for $m_h$=120 GeV.  The mass splittings are chosen to be $(\delta_1, \delta_2)=(1,1)$ GeV (left plot), (1, 10) GeV (right plot).  Red region are excluded by vacuum stability while the hatched region are excluded by perturbativity constraints.}
\label{fig:high_120}
\end{center}
\end{figure}

Fig.~\ref{fig:high_120} shows the relic density in the $m_S-\lambda_L$ plane for the high mass region: $m_S>$ 400 GeV, with $m_h$=120 GeV.  The LEP search bound is irrelevant now since the $m_S$ is much larger than the direct search limit.  The precision electroweak constraint is  weak for $m_h$ around 120 GeV as long as $\delta_1$ and $\delta_2$ do not differ too much.  The dark matter direct detection constraint is also weak due to the heavy dark matter mass.
For small mass splittings , $(\delta_1, \delta_2)=(1,1)$ GeV (left plot in Fig.~\ref{fig:high_120}) , a region of $m_S\sim500-700$ GeV  and $-0.2 < \lambda_L<0.2$ is allowed.  The dominating annihilation channels are $WW$, $WZ$ and $ZZ$.   The dependence on $\lambda_L$ is introduced by the $SSh$ coupling.
For $\lambda_L$ close to zero, a smaller value of $m_S$ is preferred to increase the annihilation cross section. When $|\lambda_L|$ gets larger, a larger value of $m_S$ is needed for the relic density to fall into the 3$\sigma$ band.
The region to the right of the 3$\sigma$ band overclose the Universe while the region to the left of the 3$\sigma$ band corresponds to the under-abundance region.  Regions with large values of $|\lambda_L|$ are excluded due to the perturbativity constraints.

The annihilation cross section grows for large mass splittings.  Therefore, the WMAP allowed region shifts to larger $m_S$ for larger $\delta_{1,2}$, as shown in the right plot of Fig.~\ref{fig:high_120} for $(\delta_1, \delta_2)=(1,10)$ GeV.
$m_S \gtrsim$ 650 GeV falls into the WMAP 3$\sigma$ region.
The result for $(\delta_1, \delta_2)=(10,1)$ GeV is very similar.  The region forbidden by perturbativity constraints (hatched region), however, shifts to the left for larger $\delta_{1,2}$.  Therefore, no allowed region is left if at least one of $\delta_{1,2} \gtrsim$ 12 GeV.

For a large SM Higgs mass $m_h=500$ GeV, a large mass splitting $\delta_1\gtrsim 150$ GeV is needed to satisfy the precision electroweak constraints.  There is no region in $m_S-\lambda_L$ that survives after  all of the experimental and theoretical constraints are taken into account.

\section{Conclusion}
\label{sec:conclusion}

We studied the simple extension of the SM Higgs sector when an extra inert Higgs doublet is introduced that couples to the gauge sector only.  The lighter of the neutral components could be a good WIMP dark matter candidate.  We explored the parameter spaces of the IHDM, taking into account the relic density constraints from WMAP and various theoretical and experimental constraints.  Table~\ref{table:summary} summarizes five distinctive regions that could provide the right amount of cold dark matter in the Universe which satisfy all of the constraints.

\begin{table}
\begin{tabular}{|c|c|c|c|c|c|}
\hline
& &$m_h$&$m_S$ (GeV)&$\lambda_L$&$\delta_1$, $\delta_2$\\ \hline
(I)&light dark matter&low $m_h$&30 $-$ 60  &$-0.15 \ {\rm to}\ 0$&50 GeV$\lesssim\delta_1\sim\delta_2\lesssim90$ GeV\\
(II)&&&60 $-$ 80&$-0.2\  {\rm to}\ 0.2$&at least one of $\delta_1$, $\delta_2$ is large\\   \cline{3-6}
(III)&&high $m_h$&50 $-$ 75&$-1 \  {\rm to}\  3$&large $\delta_1$ and small $\delta_2<$ 8 GeV\\
(IV)&&&$\sim$ 75&$-1\  {\rm to}\ 3$&large $\delta_1$ and $\delta_2$\\ \hline
(V)&heavy dark matter&low $m_h$&500$-$1000&$-0.2\  {\rm to}\ 0.3$&small $\delta_{1,2}$\\
\hline
\end{tabular}
\caption{Allowed parameter regions in the IHDM that are consistent with the WMAP dark matter relic density 3$\sigma$ region.}
\label{table:summary}
\end{table}

%\begin{itemize}

%\item{(I) Low $m_h$, $m_S \sim 20$ GeV, $\lambda_L\sim -0.2$ for large $\delta_1$ and $\delta_2$.}

%\item{(I) Low $m_h$, 60 GeV $< m_S < $ 80 GeV, $-0.2< \lambda_L <0.2$ when at least one of $\delta_1$, $\delta_2$ is large.}

%\item{(II) High $m_h$, 50 GeV $< m_S< $ 75 GeV, -1 $<\lambda_L <$ 3 for large $\delta_1$ and small $\delta_2<$ 8 GeV.}

%\item{(III) High $m_h$, $m_S\sim$ 75 GeV, -1 $<\lambda_L <$ 3 for large $\delta_1$ and $\delta_2$.}

%\item{(IV) low $m_h$, 500 GeV $<m_S<1000$ GeV, -0.2 $<\lambda_L <$ 0.2 for small $\delta_{1,2}$.}

%\end{itemize}

In regions (I) $-$ (IV), the dark matter candidate $S$, along with at least one of the other scalars $A$, $H^\pm$ is light.  Those particles could be pair produced at the Large Hadron Collider (LHC) as $SA$, $AH^\pm$, $SH^\pm$ and $H^+H^-$ with cross sections around the fb level.  Heavier scalars $A$ and $H^\pm$ could decay into the lightest one $S$ via on-shell(or off-shell) $Z^{(*)}$ and $W^{(*)}$, which further decay into quarks, leptons and neutrinos.  There are typically large missing $E_T$ in those processes due to the undetectable $S$ particles which are at the end of the decay chains. Experimentally, we can search for events with single lepton $+$ missing $E_T$, dilepton $+$ missing $E_T$, trilepton $+$ missing $E_T$ or in general, jets $+$ leptons $+$ missing $E_T$.  The dominant SM background comes from $WW$, $WZ$ and $ZZ$. The collider analysis on this model is currently under study \cite{IHDMcollider}.  The scalar mass in the high mass region (V), however, is larger than 500 GeV.  Since they only have weak interactions, the production cross section at the LHC is typically too small.

For the low $m_h$ region (II), when a relatively large $|\lambda_L |\sim 0.2$ could be accommodated, the indirect detection of the dark matter via its annihilation into neutrinos, photons, electrons and positrons could be very promising.  A recent study on the indirect neutrino signals \cite{AgrawalXZ} showed that in the low $m_S$ region, tens to a hundred of neutrino events per year from   dark matter annihilation inside the Earth and hundreds of neutrino events per year from   dark matter annihilation inside the Sun can be expected at future neutrino telescopes.
Indirect photon signals including monochromatic photon line, fragmentation photon spectrum and final state radiation photon spectrum in the IHDM is under current study \cite{IHDMphoton}.
High $m_h$ regions (III) and (IV)   typically don't have promising indirect detection signals due to the suppression of the annihilation cross section by large $m_h$.
The indirect detection possibility for the high $m_S$ region (V) is also less comparing to region (II).  In Ref.~\cite{AgrawalXZ}, it is shown that neutrino events from the Earth is too low to be observed, while a few events per year is expected from neutrinos from the sun.

In summary, the IHDM is a simple extension of the SM which provides a very promising WIMP dark matter candidate.  There are several regions of the parameter space that could provide the right amount of   dark matter relic density in the Universe.  The collider phenomenology of this model is very rich.  For certain regions, the indirect detection via neutrinos or photons is possible at future neutrino and gamma ray telescopes or ground based experiments.

\begin{acknowledgments}
We would like to thank  B.~Thomas for useful discussion and comments.
This work is supported under U.S. Department of Energy
contract\# DE-FG02-04ER-41298.

\end{acknowledgments}

\end{document}